# Measurement of nuclide cross-sections of spallation residues in 1 A GeV $^{238}$U + proton collisions

J. Taïeb[a,*], K.-H. Schmidt[b], , L. Tassan-Got[a], P. Armbruster[b], J. Benlliure[b], M. Bernas[a], A. Boudard[d], E. Casarejos[c], S. Czajkowski[e], T. Enqvist[b], R. Legrain[d†], S. Leray[d], B. Mustapha[a], M. Pravikoff[e], F. Rejmund[a], C. Stéphan[a], C. Volant[d], W. Wlazło[d]

[a]*IPN Orsay, IN2P3, 91406 Orsay, France*
[b]*GSI Darmstadt, Planckstraße 1, 64291 Darmstadt, Germany*
[c]*University of Santiago de Compostela, 15706 Santiago de Compostela, Spain*
[d]*DAPNIA/SPhN CEA Saclay, 91191 Gif sur Yvette, France*
[e]*CENBG, IN2P3, 33175 Gradignan, France*

**Abstract :** The production of heavy nuclides from the spallation-evaporation reaction of $^{238}$U induced by 1 GeV protons was studied in inverse kinematics. The evaporation residues from tungsten to uranium were identified in-flight in mass and atomic number. Their production cross-sections and their momentum distributions were determined. The data are compared with empirical systematics. A comparison with previous results from the spallation of $^{208}$Pb and $^{197}$Au reveals the strong influence of fission in the spallation of $^{238}$U.

## 1. Introduction

Since some years, spallation reactions have gained a renewed interest for several reasons. On the one hand, they are planned to be used in the so-called accelerator driven system (ADS) as an intense neutron source. On the other hand, spallation reactions lead to the production of unstable nuclei. This reaction is actually exploited in ISOL-type facilities.

Therefore, a campaign of measurements of spallation-residues started at GSI, taking advantage of the use of the inverse kinematics. The results obtained in the spallation of gold and lead have already been published extensively [1, 2, 3, 4, 5]. In both cases, the projectile energy was close or equal to 1 GeV per nucleon in order to mimic the spallation of a heavy nucleus by 1 GeV protons and 2 GeV deuterons, respectively. These measurements are supposed to give high constraints for the codes aimed for designing accelerator-driven systems and new facilities for the production of radioactive nuclei beams. They also give some clear hints for a better understanding of the spallation reaction.

---

[*] Present address : CEA/Saclay DM2S/SERMA/LENR, 91191 Gif/Yvette CEDEX, France
[†] Deceased

This paper focuses on the production of evaporation residues in the spallation of $^{238}$U. The production of those residues is conditioned by the fission of isotopes during the evaporation phase. Therefore, the fission probability estimated by the de-excitation codes can be tested by the measurement of evaporation residues. This problem is connected to some open questions on the evolution of the level density or the barrier height with increasing excitation energy. We are also able to study the dissipation in the fission process.

The measurements of evaporation residues have started since accelerators deliver relativistic protons in the early 1950s. For 40 years, evaporation-residue production was measured using chemical and/or spectroscopic methods. In 1990, at GSI the powerful heavy-ion accelerator SIS was coupled to a high-resolution recoil spectrometer, the FRS [6]. The installation of a cryogenic hydrogen target [7] permitted to start the campaign of measurements of spallation-residue cross-sections in inverse kinematics. We could detect, unambiguously identify and analyse several hundreds of primary nuclides with an accuracy in the order of 10% to 15% in most cases. This strongly contrasts with the scarce and usually cumulative cross-sections obtained with other techniques. The high efficiency of the spectrometer combined with the very short time-of-flight (about 150ns eigentime) ensures the quality of our results.

In this paper, we report on the first identification of about 365 evaporation residues, forming isotopic chains from tungsten ($Z$=74) to uranium ($Z$=92). In the second section, we present some characteristics of the experimental set up and the analysis techniques. In the third part, we report on the obtained cross-sections and kinematical properties of the studied nuclei. We discuss the results and compare them with previously established empirical systematics.

## 2. The experiment

The experimental set up was already described in other publications of experiments using gold and lead projectiles [1, 2, 3, 4, 5]. In the present chapter, we give an overview of the main aspects of the experiment and stress the improvements, which were necessary for this specific measurement.

The 1 $A$ GeV $^{238}$U beam, produced by the synchrotron SIS of GSI, interacted with a liquid hydrogen target. The thickness of the H$_2$ liquid target was determined experimentally previously [8] to be 87.3 mg/cm$^2$ (±3.0 mg/cm$^2$.) The number of incoming projectiles is recorded by a beam monitor, a secondary electron chamber [9]. The products of the reaction are separated and analysed by the recoil spectrometer FRS. The experimental apparatus is shown Figure 1. The two-stage fragment separator allows a full identification in nuclear charge $Z$ and mass number $A$ of each fragment. Moreover, the recoil momentum is also provided. The reaction products suffer a first magnetic selection; then they are slowed down in a thick passive energy degrader situated at the intermediate focal plane. A second magnetic selection is finally applied. The time-of-flight is measured between both image planes thanks to two plastic scintillation detectors. The scintillators also give a measure of the horizontal positions at the intermediate dispersive plane and at the exit. Moreover, two multiply sampling ionisation chambers (MUSIC), placed at the exit of the spectrometer, measure the energy-loss. Multi-wire proportional counters provide additional tracking information.

Therefore, for each ion passing the FRS we obtain two magnetic rigidities, a time-of-flight and an energy-loss measurement.

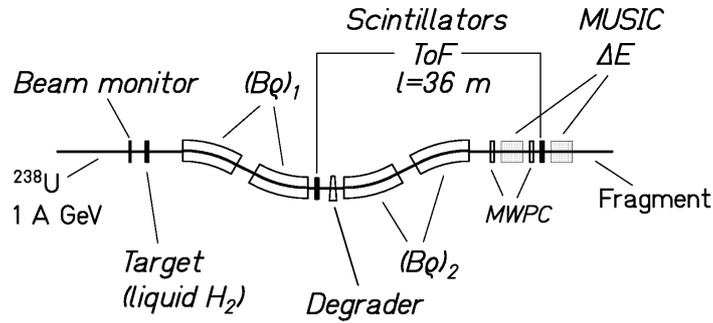

**Figure 1 : Scheme of the fragment separator (FRS) with its most important components. The primary beam of $^{238}$U enters from the left.**

**2.1. The energy loss in the degrader**

The nuclear-charge determination is certainly the most challenging problem that we had to face. It was a special aim of the experiment to improve the nuclear-charge resolution, previously obtained [10, 11]. This is especially true for the heaviest elements (the actinides) for which the separation is the most difficult. This high-resolution nuclear-charge determination could be obtained through a multi-fold measurement. First of all, we remind that a thick energy degrader is placed at the intermediate image plane (see Figure 1). This passive component of the set up indirectly helps determining the nuclear charge. Actually, the magnetic rigidities are measured before and after the ions pass through the degrader. The difference of those two quantities is linked with the momentum (and energy) loss within the thick degrader, following the relation:

$$(B\rho)_1 - (B\rho)_2 = \frac{p_1}{q_1} - \frac{p_2}{q_2} \qquad (1)$$

where $(B\rho)_1$ and $(B\rho)_2$ are the magnetic rigidities, $p_1$ and $p_2$ are the momenta, and $q_1$ and $q_2$ are the ionic charge states of the ion before and after the degrader, respectively. Assuming, for the moment that the ions are fully stripped and do not change their mass number by nuclear reactions in the degrader:

$$q_1 = q_2 = Z.e \qquad (2)$$
$$M_1 = M_2 = A\ m_0$$

the $B\rho$ difference (equation 1) provides an estimate of the energy loss within the degrader. ($m_0$ is the nuclear mass unit.) Nuclei for which the condition (2) is not fulfilled will be rejected in the analysis process as shown in the following section.

## 2.2. The nuclear-charge resolution

The best charge resolution is obtained by correlating, on the one hand, both signals coming from the ionisation chambers, and, on the other hand, the $B\rho$-difference measurement presented in the previous section. The signals provided by the ionisation chambers are combined in order to get a single optimised quantity by rejecting those events with strongly different $\Delta E$ signals in the two chambers. A bi-dimensional plot illustrating the correlation between the so-called "energy loss in the degrader" and the optimised energy loss in the MUSIC chambers is shown in Figure 2.

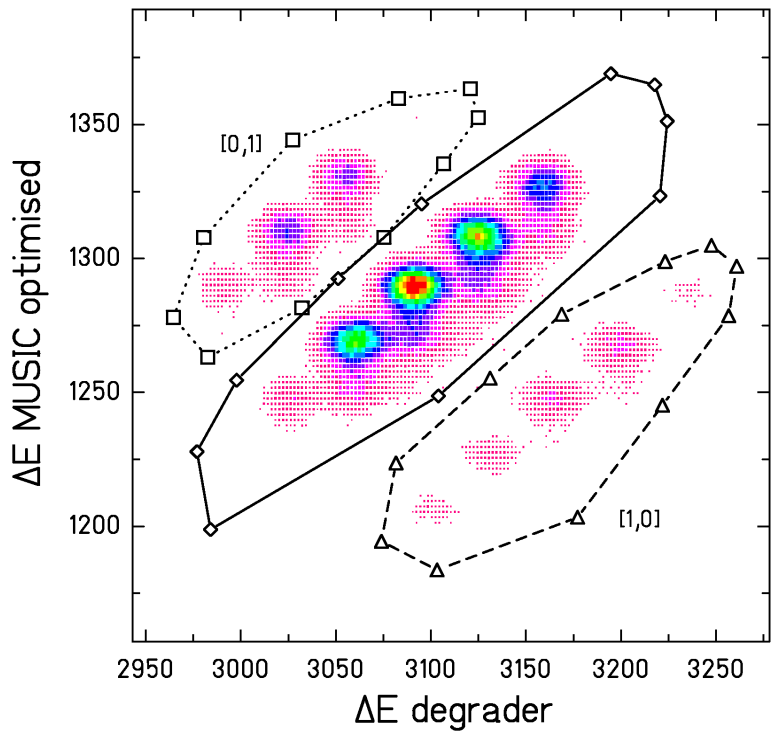

**Figure 2 : Separation of nuclear charges around Z = 90 and elimination of different ionic charge states. The ions, which do not change their charge state all along the separator, are inside the full contour line. The ions, which capture one electron in the degrader section, are inside the dotted contour line, while the ions, which loose one electron in the degrader section, are inside the dashed contour line. The most intense peak corresponds mostly to fully stripped thorium ions.**

Figure 2 is obtained for a specific setting, this means for specific values of the various magnetic fields. The number of different isotopes passing through the FRS depends on the $B\rho$ acceptance of the spectrometer and on the thickness of the energy degrader. Therefore, a number of about 70 settings was necessary to provide the full set of data presented in this paper.

Three different regions can be observed on Figure 2. They correspond to three different charge-state combinations. The central zone includes the fully stripped ions and the H-like ones. The ions in this region are bare or H-like all along their trajectory. They do not change their charge state between the first and the second half of the FRS.

Actually, most of the ions are bare after the collision at 1 GeV per nucleon. However, the probability that a heavy ion like the projectile carries on electron is not negligible; it amounts to about 10%. When arriving at S2, the ions pass through a number of different layers of matter, namely the scintillation detector and the degrader plate. Traversing those materials, the ions successively gain and lose electrons, mostly alternating between their bare and H-like state. They generally leave the intermediate image plane bare. Finally the probability that the ion is hydrogen-like all along its trajectory is rather low compared to the most probable situation that it is bare over the whole flight-path. The contamination of the central zone on Figure 2 due to the ions carrying one electron in *both* sections of the FRS is estimated to be in the order of 1% to 2%, depending on the atomic number of the ion [12]. The higher it is, the higher is the contamination. This contamination is neglected in the analysis.

The two other zones (labelled [1,0] and [0,1]) are to be associated to the ions carrying one electron in the first (region [1,0]) or second (region [0,1]) section of the FRS, being bare in the other part of the spectrometer. Only the central region in Figure 2 is being analysed for getting the cross-sections. Neglecting the contamination due to ions, which carry one electron in both sections, we assume, gating on the central region, that all ions are fully stripped. Therefore, the condition (2) is valid.

Each spot within the selected region corresponds to a common value of the energy loss in the degrader and in the MUSIC chambers. This correlation is the best way for disentangling the various elements traversing the spectrometer. Thus, each spot corresponds to a specific element. The separation is seen to be rather good. Projecting the central window onto an inclined axis, we obtain a curve whose peak-to-valley ratio varies between 10 and 20. It is the first time that such a high nuclear-charge resolution could be obtained in an in-flight-separation experiment, exploring elements up to uranium.

After selection of a specific spot and thus a specific nuclear charge in the central region of Figure 2, the mass spectrum is obtained thanks to the $B\rho$ and velocity measurements in the second section of the FRS according to the following expression.

$$A = \frac{Z \cdot e \cdot (B\rho)_2}{m_0 \cdot c \cdot \beta_2 \gamma_2} \qquad (3)$$

where, $\beta_2$ and $\gamma_2$ are the reduced velocity and the Lorentz parameter, respectively, in the second part of the spectrometer. They are deduced from the time-of-flight (ToF) measurement. $Z$ is the nuclear charge, and $e$, $m_0$ and $c$ are the charge of the electron, the mass unit, and the velocity of light, respectively. The following two-dimensional plot (Figure 3) of the mass versus the position at the intermediate image plane of nuclei around $^{192}$Pb illustrates the high mass resolution. This precise mass measurement is achieved due to the high ToF resolution (130 ps) and a long flight path (36 m). The resulting mass resolution is

$$A/\Delta A = 440 \text{ (FWHM)} \qquad (4)$$

Figure 3 shows that many isotopes are cut at the intermediate image plane due to the limited $B\rho$ acceptance of the spectrometer. Therefore, 70 settings of the fields are necessary for covering the whole range in magnetic rigidity. The broadening of the horizontal distribution at S2 reflects the extension of the velocity distribution mainly due to the nuclear reaction.

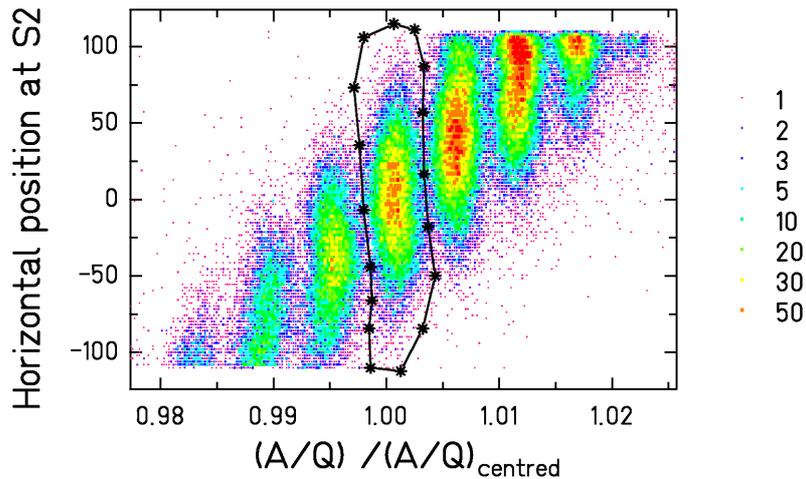

**Figure 3 : Two-dimensional cluster plot of the horizontal position at the central image plane (S2) versus the A/Q value, normalized to the one of the centred nucleus. The data are recorded in one specific setting of the fragment separator. The contour line indicates the centred nucleus, $^{192}$Pb. The colour code gives the counts per channel.**

The production rate could be obtained for 365 nuclei ranging from uranium to tungsten. For getting cross-sections, losses due to non-fully stripped ions and nuclear reactions at S2 are accounted for. The losses are estimated theoretically, and these evaluations are confirmed with online measurements. The contribution to the production rate from the windows of the target was measured during the beam time using an empty target. This part is measured to lie between 5% and 15% of the total production rate. This contribution was corrected for. More details about the applied correction procedure are given in [4].

The purpose of this work was to cover all notably produced evaporation residues. Actually, the region around $Z = 60$ to $Z = 70$ is only slightly populated, being situated in the low-mass tail of the spallation-evaporation and in the high-mass tail of the spallation-fission areas. In this region, we could not disentangle the fission fragments from the evaporation residues. Therefore, we chose not to give data for this part and restrict ourselves to the nuclei identified as pure evaporation residues ($Z > 73$). Also the contributions from secondary reactions in the hydrogen target become more important for the lighter elements, as discussed below.

In addition to the production cross-sections, we also measured, for each of these nuclei, the recoil velocity distribution characterised by its mean value and its standard deviation. Those data are presented in section 3.4.

## 3. Results and discussion

### 3.1 Measured cross-sections

Figure 4 shows the measured isotopic cross-sections for all elements ranging from uranium (Z=92) to tungsten (Z=74). They vary from about 100 mb to 10 µb. The numerical values are listed in table 1. The data represent the cross-sections obtained in a target of 87.3 mg/cm$^2$ hydrogen. The attenuation of the beam, which amounts to about 10 % along the whole target, is considered. While the contribution of the target-container windows is subtracted, the production rates are only partly corrected for secondary reactions inside the production target. It is assumed that secondary reactions, which are estimated on the basis of the total reaction probability after the first reaction, depopulate the nuclides in the considered range, whereas the population due to secondary reactions is assumed to appear in lighter nuclei with Z < 74 and could be neglected. We applied a correction of 5 %. To perform a more realistic correction would require a realistic calculation of all nuclide formation cross-sections starting from each of the reaction products. Such a code is not available in the moment. However, due to our experience on this procedure applied to the $^{208}$Pb +$^{1}$H system [4], this correction is expected to remain well inside the uncertainties given in this work, being most important for the lightest nuclei.

The fluctuations within the isotopic distributions are rather low. However, a few dips can be observed for elements from thorium to astatine (see Figure 4). Actually, the very short time-of-flight of the ions, 150 ns eigen-time, authorises the measurement of the production cross-sections for most of the isotopes, namely when the radioactive-decay period is much longer than the time-of-flight of the ions. Only a few isotopes characterized by a number of 128 neutrons decay by α emission towards the 126-neutron shell and cause a dip in the apparent cross-sections. The decay period of those isotopes is of the order of the time-of-flight through the FRS. Therefore, part of the production is lost before being analysed and identified. Moreover, when the decay occurs at the very beginning of the flight path, the ion is identified as the daughter nucleus. This effect causes the slight hump that can be observed for $^{211}$At. The apparent over production of this isotope is due to the very fast decay of $^{215}$Fr, which has a half-life of $T_\alpha$=90 ns to be compared to the ToF=150 ns eigen-time. It is not simple to correct for this effect, since many of these nuclei might be produced in isomeric states. Moreover, the branching ratios and some of the associated decay periods are not known. A couple of isotones with a number of 129 neutrons are also concerned. Mainly $^{214}$At suffers from the very fast decay. The very fast decay of these nuclei, which prevents us from giving true cross-sections in these few cases, demonstrates the high quality of our data. This ensures that the uncertainty is as low as claimed, since effects of 10% to 20% can be observed. Moreover, we guarantee that the mass and charge identification is correct.

We observed, for the first time, the isotope $^{235}$Ac that corresponds to the 3-proton removal channel. 150 events were unambiguously recorded. This points out again [13] that the so-called cold fragmentation is a favourable path for producing heavy exotic neutron-rich nuclei.

Thanks to the measurement of the full isotopic chains, we are able to produce sensible isobaric cross-sections, this means, cross-sections summed up over the full isobaric chains. The associated plot is reproduced in Figure 5, which includes two other distributions. Thus, we show, on the same graph, the isobaric cross-sections obtained by the same collaboration, following a similar procedure, with $^{197}$Au projectiles at 800 MeV per nucleon [1,8] and $^{208}$Pb projectiles at 1 GeV per nucleon [3, 4]. The isobaric cross-section is shown in Figure 5 as a function of the mass loss. Summing up the measured nuclide cross-sections, we obtain a total spallation-evaporation cross-section of 420 mb. The extrapolation of the mass distribution to lighter masses not covered in the experiment, results in an estimated total spallation-evaporation cross-section of 460 mb.

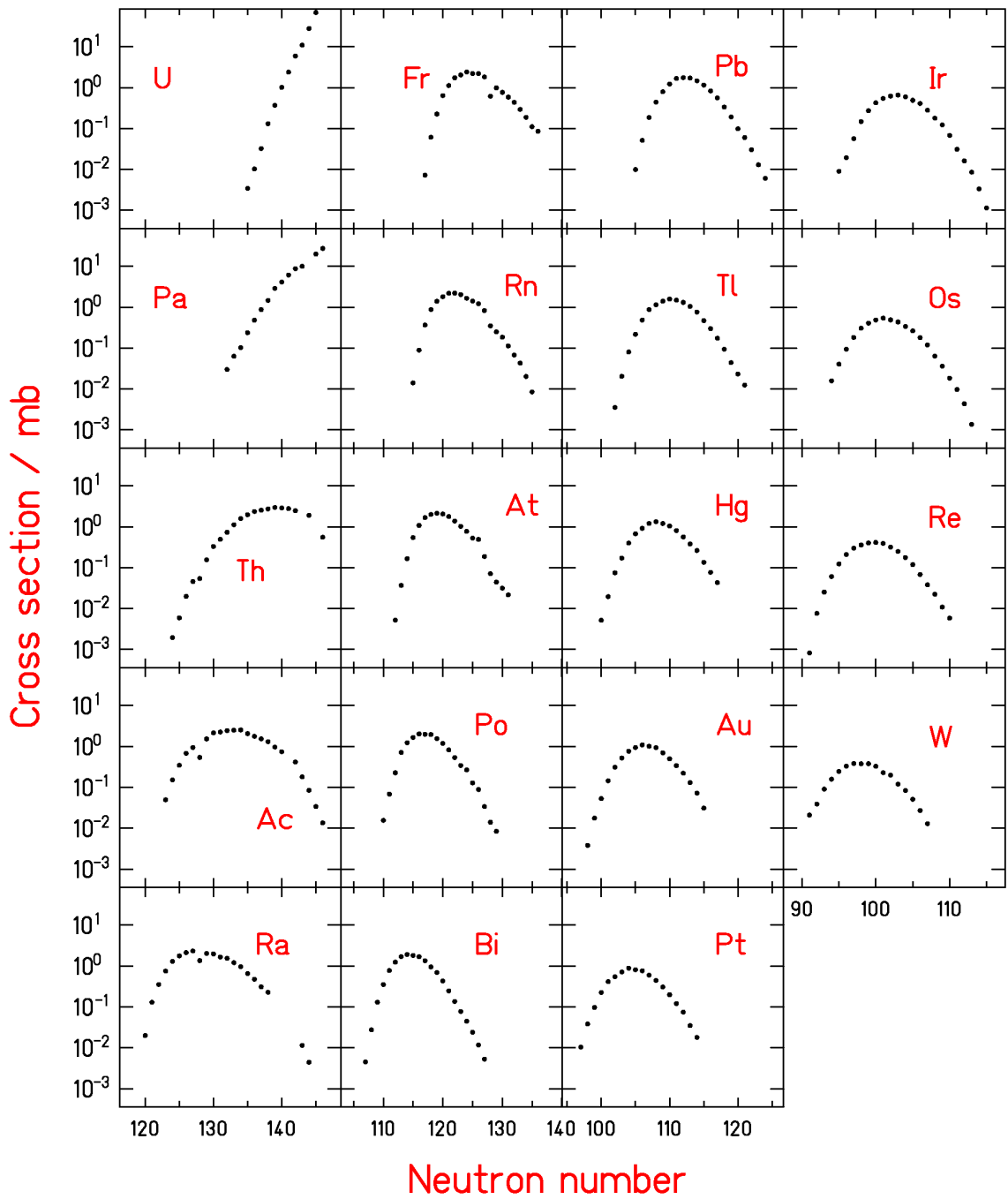

**Figure 4 :** Isotopic production cross-sections for 365 nuclides from the spallation-evaporation reaction of $^{238}$U by 1 GeV protons. The non-systematic uncertainties are smaller than the point size. The measurement was performed in inverse kinematics.

**Table 1:** Numerical values of the isotopic production cross-sections for 365 nuclides from the spallation-evaporation reaction of $^{238}$U by 1 GeV protons. The measurement was performed in inverse kinematics. The numbers give the nuclear charge Z, the mass number A, the measured cross-section and the systematic (unc1) and the total (unc2) relative uncertainty in percent. The values given in brackets are apparent production cross-sections, influenced by the radioactive decay inside the spectrometer. In these cases, no uncertainty is given.

| Z | A | σ/mb | unc.1 | unc. 2 | Z | A | σ/mb | unc.1 | unc. 2 | Z | A | σ/mb | unc.1 | unc. 2 |
|---|---|---|---|---|---|---|---|---|---|---|---|---|---|---|
| 74 | 165 | 0.019 | 12 | 15 | 76 | 176 | 0.49 | 11 | 14 | 78 | 184 | 0.76 | 10 | 12 |
| 74 | 166 | 0.035 | 12 | 15 | 76 | 177 | 0.54 | 11 | 14 | 78 | 185 | 0.60 | 10 | 12 |
| 74 | 167 | 0.082 | 12 | 15 | 76 | 178 | 0.49 | 11 | 14 | 78 | 186 | 0.44 | 9 | 12 |
| 74 | 168 | 0.14 | 12 | 15 | 76 | 179 | 0.44 | 11 | 14 | 78 | 187 | 0.31 | 9 | 12 |
| 74 | 169 | 0.22 | 12 | 15 | 76 | 180 | 0.34 | 11 | 14 | 78 | 188 | 0.20 | 9 | 12 |
| 74 | 170 | 0.30 | 12 | 15 | 76 | 181 | 0.26 | 11 | 14 | 78 | 189 | 0.12 | 9 | 12 |
| 74 | 171 | 0.35 | 12 | 15 | 76 | 182 | 0.18 | 10 | 14 | 78 | 190 | 0.075 | 8 | 12 |
| 74 | 172 | 0.34 | 12 | 15 | 76 | 183 | 0.12 | 10 | 14 | 78 | 191 | 0.035 | 8 | 12 |
| 74 | 173 | 0.34 | 12 | 15 | 76 | 184 | 0.064 | 10 | 14 | 78 | 192 | 0.018 | 8 | 12 |
| 74 | 174 | 0.30 | 12 | 15 | 76 | 185 | 0.036 | 10 | 14 | 79 | 177 | 0.0038 | 10 | 11 |
| 74 | 175 | 0.21 | 12 | 15 | 76 | 186 | 0.018 | 9 | 14 | 79 | 178 | 0.018 | 10 | 11 |
| 74 | 176 | 0.18 | 11 | 15 | 76 | 187 | 0.0097 | 9 | 14 | 79 | 179 | 0.053 | 10 | 11 |
| 74 | 177 | 0.11 | 11 | 15 | 76 | 188 | 0.0044 | 9 | 14 | 79 | 180 | 0.14 | 10 | 11 |
| 74 | 178 | 0.076 | 11 | 15 | 76 | 189 | 0.0014 | 9 | 14 | 79 | 181 | 0.31 | 10 | 11 |
| 74 | 179 | 0.046 | 11 | 15 | 77 | 172 | 0.0090 | 12 | 13 | 79 | 182 | 0.52 | 10 | 11 |
| 74 | 180 | 0.025 | 11 | 15 | 77 | 173 | 0.019 | 12 | 13 | 79 | 183 | 0.76 | 10 | 11 |
| 74 | 181 | 0.012 | 11 | 15 | 77 | 174 | 0.057 | 12 | 13 | 79 | 184 | 0.94 | 10 | 11 |
| 75 | 166 | 0.0012 | 12 | 15 | 77 | 175 | 0.15 | 12 | 13 | 79 | 185 | 1.09 | 10 | 11 |
| 75 | 167 | 0.0076 | 12 | 15 | 77 | 176 | 0.27 | 11 | 13 | 79 | 186 | 1.01 | 10 | 11 |
| 75 | 168 | 0.025 | 12 | 15 | 77 | 177 | 0.43 | 11 | 13 | 79 | 187 | 0.93 | 9 | 11 |
| 75 | 169 | 0.061 | 12 | 15 | 77 | 178 | 0.54 | 11 | 13 | 79 | 188 | 0.70 | 9 | 11 |
| 75 | 170 | 0.12 | 12 | 15 | 77 | 179 | 0.62 | 11 | 13 | 79 | 189 | 0.50 | 9 | 11 |
| 75 | 171 | 0.21 | 12 | 15 | 77 | 180 | 0.66 | 11 | 13 | 79 | 190 | 0.34 | 8 | 11 |
| 75 | 172 | 0.30 | 12 | 15 | 77 | 181 | 0.59 | 11 | 13 | 79 | 191 | 0.22 | 8 | 11 |
| 75 | 173 | 0.36 | 12 | 15 | 77 | 182 | 0.49 | 10 | 13 | 79 | 192 | 0.13 | 8 | 11 |
| 75 | 174 | 0.40 | 12 | 15 | 77 | 183 | 0.41 | 10 | 13 | 79 | 193 | 0.072 | 8 | 11 |
| 75 | 175 | 0.42 | 12 | 15 | 77 | 184 | 0.28 | 10 | 13 | 79 | 194 | 0.031 | 8 | 11 |
| 75 | 176 | 0.39 | 11 | 15 | 77 | 185 | 0.18 | 10 | 13 | 80 | 180 | 0.0052 | 9 | 10 |
| 75 | 177 | 0.32 | 11 | 15 | 77 | 186 | 0.12 | 9 | 13 | 80 | 181 | 0.020 | 9 | 10 |
| 75 | 178 | 0.25 | 11 | 15 | 77 | 187 | 0.068 | 9 | 13 | 80 | 182 | 0.075 | 9 | 10 |
| 75 | 179 | 0.18 | 11 | 15 | 77 | 188 | 0.031 | 9 | 13 | 80 | 183 | 0.17 | 9 | 10 |
| 75 | 180 | 0.12 | 11 | 15 | 77 | 189 | 0.016 | 9 | 13 | 80 | 184 | 0.40 | 9 | 10 |
| 75 | 181 | 0.068 | 11 | 15 | 77 | 190 | 0.0085 | 8 | 13 | 80 | 185 | 0.67 | 9 | 10 |
| 75 | 182 | 0.038 | 10 | 15 | 77 | 191 | 0.0033 | 8 | 13 | 80 | 186 | 0.92 | 9 | 10 |
| 75 | 183 | 0.022 | 10 | 15 | 77 | 192 | 0.0011 | 8 | 13 | 80 | 187 | 1.21 | 8 | 10 |
| 75 | 184 | 0.011 | 10 | 15 | 78 | 175 | 0.010 | 12 | 12 | 80 | 188 | 1.33 | 8 | 10 |
| 75 | 185 | 0.0058 | 10 | 15 | 78 | 176 | 0.038 | 11 | 12 | 80 | 189 | 1.20 | 8 | 10 |
| 76 | 170 | 0.016 | 12 | 14 | 78 | 177 | 0.097 | 11 | 12 | 80 | 190 | 1.04 | 8 | 10 |
| 76 | 171 | 0.041 | 12 | 14 | 78 | 178 | 0.23 | 11 | 12 | 80 | 191 | 0.81 | 8 | 10 |
| 76 | 172 | 0.094 | 12 | 14 | 78 | 179 | 0.42 | 11 | 12 | 80 | 192 | 0.57 | 8 | 10 |
| 76 | 173 | 0.18 | 12 | 14 | 78 | 180 | 0.55 | 11 | 12 | 80 | 193 | 0.38 | 7 | 10 |
| 76 | 174 | 0.31 | 12 | 14 | 78 | 181 | 0.71 | 11 | 12 | 80 | 194 | 0.26 | 7 | 10 |
| 76 | 175 | 0.41 | 12 | 14 | 78 | 182 | 0.89 | 10 | 12 | 80 | 195 | 0.14 | 7 | 10 |
| | | | | | 78 | 183 | 0.81 | 10 | 12 | 80 | 196 | 0.077 | 7 | 10 |
| 80 | 197 | 0.043 | 7 | 10 | 83 | 201 | 0.95 | 7 | 11 | 86 | 203 | 0.37 | 8 | 13 |
| 81 | 183 | 0.0035 | 9 | 10 | 83 | 202 | 0.69 | 7 | 11 | 86 | 204 | 0.88 | 8 | 13 |
| 81 | 184 | 0.020 | 9 | 10 | 83 | 203 | 0.43 | 7 | 11 | 86 | 205 | 1.40 | 8 | 13 |
| 81 | 185 | 0.081 | 9 | 10 | 83 | 204 | 0.25 | 7 | 11 | 86 | 206 | 1.81 | 8 | 13 |
| 81 | 186 | 0.22 | 9 | 10 | 83 | 205 | 0.14 | 7 | 11 | 86 | 207 | 2.19 | 8 | 13 |
| 81 | 187 | 0.49 | 8 | 10 | 83 | 206 | 0.077 | 7 | 11 | 86 | 208 | 2.21 | 8 | 13 |
| 81 | 188 | 0.87 | 8 | 10 | 83 | 207 | 0.045 | 7 | 11 | 86 | 209 | 2.03 | 8 | 13 |
| 81 | 189 | 1.15 | 8 | 10 | 83 | 208 | 0.024 | 7 | 11 | 86 | 210 | 1.66 | 8 | 13 |
| 81 | 190 | 1.41 | 8 | 10 | 83 | 209 | 0.012 | 7 | 11 | 86 | 211 | 1.41 | 8 | 13 |
| 81 | 191 | 1.59 | 8 | 10 | 83 | 210 | 0.0053 | 7 | 11 | 86 | 212 | [1.23] | | |
| 81 | 192 | 1.49 | 8 | 10 | 84 | 194 | 0.016 | 7 | 12 | 86 | 213 | 0.83 | 8 | 13 |
| 81 | 193 | 1.31 | 7 | 10 | 84 | 195 | 0.068 | 7 | 12 | 86 | 214 | [0.35] | | |
| 81 | 194 | 1.04 | 7 | 10 | 84 | 196 | 0.23 | 7 | 12 | 86 | 215 | [0.25] | | |
| 81 | 195 | 0.76 | 7 | 10 | 84 | 197 | 0.71 | 7 | 12 | 86 | 216 | 0.19 | 8 | 13 |
| 81 | 196 | 0.47 | 7 | 10 | 84 | 198 | 1.23 | 7 | 12 | 86 | 217 | 0.11 | 8 | 13 |
| 81 | 197 | 0.30 | 7 | 10 | 84 | 199 | 1.66 | 7 | 12 | 86 | 218 | 0.068 | 8 | 13 |
| 81 | 198 | 0.17 | 7 | 10 | 84 | 200 | 2.02 | 7 | 12 | 86 | 219 | 0.043 | 8 | 13 |
| 81 | 199 | 0.094 | 7 | 10 | 84 | 201 | 1.98 | 7 | 12 | 86 | 220 | 0.020 | 8 | 13 |
| 81 | 200 | 0.044 | 7 | 10 | 84 | 202 | 1.95 | 7 | 12 | 86 | 221 | 0.0085 | 8 | 13 |
| 81 | 201 | 0.023 | 7 | 10 | 84 | 203 | 1.55 | 7 | 12 | 87 | 204 | 0.0072 | 8 | 14 |
| 81 | 202 | 0.012 | 7 | 10 | 84 | 204 | 1.19 | 7 | 12 | 87 | 205 | 0.061 | 8 | 14 |
| 82 | 187 | 0.0099 | 7 | 10 | 84 | 205 | 0.82 | 7 | 12 | 87 | 206 | 0.23 | 8 | 14 |
| 82 | 188 | 0.051 | 7 | 10 | 84 | 206 | 0.54 | 7 | 12 | 87 | 207 | 0.64 | 8 | 14 |
| 82 | 189 | 0.18 | 7 | 10 | 84 | 207 | 0.34 | 7 | 12 | 87 | 208 | 1.13 | 8 | 14 |
| 82 | 190 | 0.44 | 7 | 10 | 84 | 208 | 0.27 | 7 | 12 | 87 | 209 | 1.75 | 8 | 14 |
| 82 | 191 | 0.79 | 7 | 10 | 84 | 209 | 0.13 | 7 | 12 | 87 | 210 | 2.05 | 8 | 14 |
| 82 | 192 | 1.23 | 7 | 10 | 84 | 210 | [0.088] | | | 87 | 211 | 2.41 | 8 | 14 |
| 82 | 193 | 1.68 | 7 | 10 | 84 | 211 | 0.034 | 7 | 12 | 87 | 212 | 2.22 | 8 | 14 |
| 82 | 194 | 1.76 | 7 | 10 | 84 | 212 | [0.014] | | | 87 | 213 | [2.20] | | |
| 82 | 195 | 1.72 | 7 | 10 | 84 | 213 | [0.0084] | | | 87 | 214 | 1.82 | 8 | 14 |
| 82 | 196 | 1.47 | 7 | 10 | 85 | 197 | 0.0052 | 8 | 13 | 87 | 215 | [0.62] | | |
| 82 | 197 | 1.16 | 7 | 10 | 85 | 198 | 0.037 | 8 | 13 | 87 | 216 | [0.99] | | |
| 82 | 198 | 0.83 | 7 | 10 | 85 | 199 | 0.17 | 8 | 13 | 87 | 217 | 0.77 | 8 | 14 |
| 82 | 199 | 0.56 | 7 | 10 | 85 | 200 | 0.55 | 8 | 13 | 87 | 218 | 0.59 | 8 | 14 |
| 82 | 200 | 0.33 | 7 | 10 | 85 | 201 | 1.08 | 8 | 13 | 87 | 219 | 0.44 | 8 | 14 |
| 82 | 201 | 0.19 | 7 | 10 | 85 | 202 | 1.68 | 8 | 13 | 87 | 220 | 0.29 | 8 | 14 |
| 82 | 202 | 0.098 | 7 | 10 | 85 | 203 | 2.01 | 8 | 13 | 87 | 221 | 0.19 | 8 | 14 |
| 82 | 203 | 0.060 | 7 | 10 | 85 | 204 | 2.15 | 8 | 13 | 87 | 222 | 0.11 | 8 | 14 |
| 82 | 204 | 0.030 | 7 | 10 | 85 | 205 | 2.06 | 8 | 13 | 87 | 223 | 0.085 | 8 | 14 |
| 82 | 205 | 0.013 | 7 | 10 | 85 | 206 | 1.79 | 8 | 13 | 88 | 208 | 0.020 | 8 | 14 |
| 82 | 206 | 0.0061 | 7 | 10 | 85 | 207 | 1.39 | 8 | 13 | 88 | 209 | 0.13 | 8 | 14 |
| 83 | 190 | 0.0045 | 7 | 11 | 85 | 208 | 1.03 | 8 | 13 | 88 | 210 | 0.35 | 8 | 14 |
| 83 | 191 | 0.027 | 7 | 11 | 85 | 209 | 0.78 | 8 | 13 | 88 | 211 | 0.75 | 8 | 14 |
| 83 | 192 | 0.13 | 7 | 11 | 85 | 210 | 0.53 | 8 | 13 | 88 | 212 | 1.31 | 8 | 14 |
| 83 | 193 | 0.35 | 7 | 11 | 85 | 211 | [0.49] | | | 88 | 213 | 1.77 | 8 | 14 |
| 83 | 194 | 0.78 | 7 | 11 | 85 | 212 | 0.19 | 8 | 13 | 88 | 214 | 2.13 | 8 | 14 |
| 83 | 195 | 1.25 | 7 | 11 | 85 | 213 | [0.071] | | | 88 | 215 | 2.34 | 8 | 14 |
| 83 | 196 | 1.70 | 7 | 11 | 85 | 214 | [0.044] | | | 88 | 216 | [1.36] | | |
| 83 | 197 | 1.92 | 7 | 11 | 85 | 215 | 0.032 | 8 | 13 | 88 | 217 | [2.05] | | |
| 83 | 198 | 1.82 | 7 | 11 | 85 | 216 | 0.022 | 8 | 13 | 88 | 218 | 1.98 | 8 | 14 |
| 83 | 199 | 1.69 | 7 | 11 | 86 | 201 | 0.014 | 8 | 13 | 88 | 219 | 1.66 | 8 | 14 |
| 83 | 200 | 1.35 | 7 | 11 | 86 | 202 | 0.089 | 8 | 13 | 88 | 220 | 1.55 | 8 | 14 |
| 88 | 221 | 1.21 | 8 | 14 | 89 | 232 | 0.18 | 8 | 15 | 91 | 225 | 0.10 | 8 | 15 |
| 88 | 222 | 0.96 | 8 | 14 | 89 | 233 | 0.085 | 8 | 15 | 91 | 226 | 0.24 | 8 | 15 |
| 88 | 223 | 0.66 | 8 | 14 | 89 | 234 | 0.034 | 8 | 15 | 91 | 227 | 0.48 | 8 | 15 |
| 88 | 224 | 0.48 | 8 | 14 | 89 | 235 | 0.013 | 8 | 16 | 91 | 228 | 0.88 | 8 | 15 |
| 88 | 225 | 0.31 | 8 | 14 | 90 | 214 | 0.0020 | 8 | 15 | 91 | 229 | 1.47 | 8 | 15 |
| 88 | 226 | 0.23 | 8 | 14 | 90 | 215 | 0.0059 | 8 | 15 | 91 | 230 | 2.88 | 8 | 15 |
| 88 | 231 | 0.011 | 8 | 14 | 90 | 216 | 0.020 | 8 | 15 | 91 | 231 | 4.12 | 8 | 15 |
| 88 | 232 | 0.0044 | 8 | 14 | 90 | 217 | 0.046 | 8 | 15 | 91 | 232 | 6.13 | 8 | 15 |
| 89 | 212 | 0.050 | 8 | 15 | 90 | 218 | [0.054] | | | 91 | 233 | 8.81 | 8 | 15 |
| 89 | 213 | 0.15 | 8 | 15 | 90 | 219 | [0.16] | | | 91 | 234 | 10.0 | 8 | 15 |
| 89 | 214 | 0.35 | 8 | 15 | 90 | 220 | 0.33 | 8 | 15 | 91 | 236 | 19.9 | 8 | 15 |
| 89 | 215 | 0.68 | 8 | 15 | 90 | 221 | 0.50 | 8 | 15 | 91 | 237 | 27.5 | 8 | 15 |
| 89 | 216 | 0.94 | 8 | 15 | 90 | 222 | 0.74 | 8 | 15 | 92 | 227 | 0.0034 | 8 | 15 |
| 89 | 217 | [0.54] | | | 90 | 223 | 1.12 | 8 | 15 | 92 | 228 | 0.010 | 8 | 15 |
| 89 | 218 | [1.53] | | | 90 | 224 | 1.59 | 8 | 15 | 92 | 229 | 0.032 | 8 | 15 |
| 89 | 219 | 2.15 | 8 | 15 | 90 | 225 | 1.99 | 8 | 15 | 92 | 230 | 0.13 | 8 | 15 |
| 89 | 220 | 2.25 | 8 | 15 | 90 | 226 | 2.37 | 8 | 15 | 92 | 231 | 0.37 | 8 | 15 |
| 89 | 221 | 2.43 | 8 | 15 | 90 | 227 | 2.56 | 8 | 15 | 92 | 232 | 1.01 | 8 | 15 |
| 89 | 222 | 2.49 | 8 | 15 | 90 | 228 | 2.74 | 8 | 15 | 92 | 233 | 2.40 | 8 | 15 |
| 89 | 223 | 2.54 | 8 | 15 | 90 | 229 | 2.97 | 8 | 15 | 92 | 234 | 5.95 | 8 | 15 |
| 89 | 224 | 2.06 | 8 | 15 | 90 | 230 | 2.90 | 8 | 15 | 92 | 235 | 11.0 | 8 | 15 |
| 89 | 225 | 1.77 | 8 | 15 | 90 | 231 | 2.78 | 8 | 15 | 92 | 236 | 28.1 | 8 | 15 |
| 89 | 226 | 1.53 | 8 | 15 | 90 | 232 | 2.48 | 8 | 15 | 92 | 237 | 68.7 | 8 | 15 |
| 89 | 227 | 1.31 | 8 | 15 | 90 | 234 | 1.92 | 8 | 15 | | | | | |
| 89 | 228 | 0.97 | 8 | 15 | 90 | 236 | 0.56 | 8 | 15 | | | | | |
| 89 | 229 | 0.74 | 8 | 15 | 91 | 223 | 0.030 | 8 | 15 | | | | | |
| 89 | 231 | 0.42 | 8 | 15 | 91 | 224 | 0.063 | 8 | 15 | | | | | |

As far as nuclei with low fissility as gold and lead are concerned, the trends are very much similar. The largest differences can be observed for the lowest masses. The lighter evaporation residues are less produced in the gold experiment. The explanation appears clearly when considering the difference in projectile energy. Within the frame of a two-stage model of the spallation reaction [14], the first phase of the interaction leads to the production of an excited nucleus. The excitation spectrum depends on the energy of the projectile. The faster projectile leads to higher excitation energy, as far as the so-called limiting fragmentation regime is not reached. This leads to longer evaporation chains and higher production of the lighter evaporation residues. The height of both distributions from the spallation of gold and lead is similar for the heaviest residues. The shape and the height of the isobaric distributions resulting from the spallation of uranium are very different.

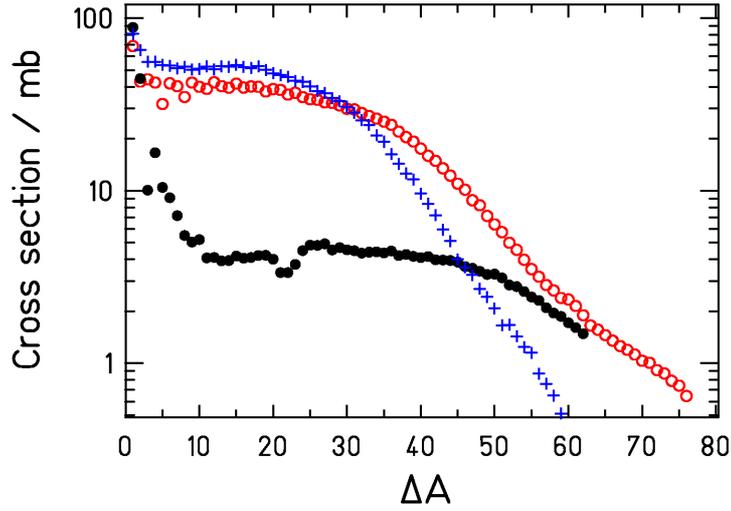

**Figure 5 :** Isobaric cross-sections as a function of mass loss for three reactions. The full symbols mark the system $^{238}$U + $^{1}$H at 1 $A$ GeV studied in this work, the open symbols represent the system $^{208}$Pb + $^{1}$H at 1 $A$ GeV [4], and the crosses result from the reaction $^{197}$Au + $^{1}$H at 800 $A$ MeV [1]. The non-systematic uncertainties are smaller than the point size.

In the isobaric cross-sections of the uranium system, we notice a slight dip for mass losses close to 22. This depression is due to the very fast alpha decay of $N$ = 128 isotones towards the 126-neutron shell described above. Moreover, for mass losses smaller than 60 units, the cross-sections are notably lower in the uranium case than obtained for the two other experiments. The effect is stronger for the heaviest fragments. The explanation lies in the strong depletion effect by the fission process for actinide nuclei. During the evaporation phase, the fission probability is much higher for actinides ($Z$ > 88) than for lighter elements, which are involved in the spallation of gold or lead, respectively. Thus, in the case of uranium, the production of evaporation residues is strongly influenced by the fission mechanism. This observation goes in line with previous results in the fragmentation of $^{238}$U and $^{208}$Pb in a copper target [15]. This observation is convincingly demonstrated comparing the isotopic cross-section for production of projectile isotopes in the uranium, lead and gold experiments. The cross-sections are plotted in Figure 6 as a function of the neutron loss for both experiments. The distributions appear very different. The cross-sections are similar for the one-neutron removal channel, but they rapidly diverge when increasing the number of lost neutrons. A longer de-excitation chain produces the more neutron-deficient isotopes. Two effects contribute to the fast decrease of the production cross-section with increasing neutron loss. First-of-all, the longer is the de-excitation chain the higher is the cumulated fission probability. Obviously, longer evaporation chains produce the lighter uranium isotopes. Moreover, the fission barriers decrease for the more neutron-deficient nuclei, favouring again the fission process relative to the evaporation of particles. Thus, as the neutron-rich part of the isotopic distribution is not so much affected by the fission mechanism, the neutron-deficient region is strongly depopulated. Close to the projectile, the length of the isotopic chains is notably shorter in the uranium case than in the gold one. It is an interesting observation that in the same experiment we produce low and highly excited very fissile pre-fragments. Therefore, the data provide relevant information on the competition between fission and evaporation in a wide range of fissility and excitation.

Another qualitative observation deduced from Figure 5 is that the isobaric distributions associated to the lead and uranium experiments join for mass loss close to 60. The depletion effect observed on the cross-section for the heaviest nuclei seems to vanish for the lightest evaporation residues. Coming back to the two-stage model, this observation seems surprising. Actually, the first step of the reaction leads to the production of an excited so-called pre-fragment. The mass and nuclear charge of this nucleus is close to the one of the projectile, only 5 to 10 nucleons could be removed [16]. Therefore, the light evaporation residues, for instance with a mass loss equal to 60 compared to the projectile, are produced in very long evaporation chains starting in the actinide region. The light evaporation residues are not depopulated in the uranium case compared to the lead, which indicates that the fission probability along the extensive de-excitation path is rather low, in spite of the fact that this path crosses the actinide and pre-actinide regions, where the fission barriers are low. The explanation lies in the inhibition of the fission process for highly excited nuclei. Indeed, nuclear fission is a dynamical process, which needs time to develop. In a macroscopic picture of the nucleus, Grangé and Weidenmüller [17], following the pioneering work of Kramers [18], treated fission as a diffusive process over the potential barrier, which is governed by nuclear viscosity. Experimental information on the magnitude of nuclear viscosity is still controversial. Therefore, the residue cross-sections, determined in this work, provide valuable information on this fundamental nuclear property. A dedicated quantitative analysis will be presented in a separate publication.

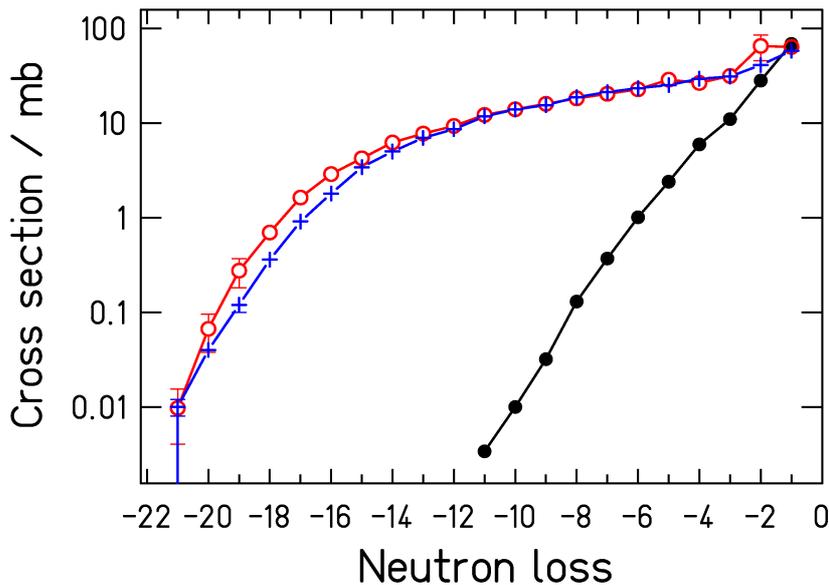

**Figure 6 : Isotopic cross-section for production of projectile isotopes in the uranium (full symbols), lead (open symbols) [1], and gold (crosses) [4] experiments. The cross-sections are plotted as a function of the neutron loss.**

### 3.2. Comparison with other data

We could find very few results, which can be compared to ours. Actually, three different experiments matched to the present one. Lindner and Osborne [19] studied the evaporation-residue production after the spallation of $^{238}$U by 340 MeV protons, Pate and Poskanzer [20] studied the same reaction at 680 MeV. The projectile energy is sensibly lower but we expect little variations due to the projectile energy modification for the heaviest fragments. They irradiated the uranium target and applied chemical techniques followed by a spectroscopic analysis. Titarenko and collaborators [21, 22] also measured several evaporation residues produced by the same reaction at 800 MeV. They used pure spectroscopic methods. This experimental technique is best suited for determining cumulative yields of radioactive decay chains. It also allows the measurement of independent yields of shielded nuclei. These are nuclei not produced by the decay of any potential mother isotopes.

Figure 7 compares the previously measured cross-sections obtained by Lindner and Osborne [19], by Pate and Poskanzer [20], and by Titarenko and collaborators [21, 22] to the present data.

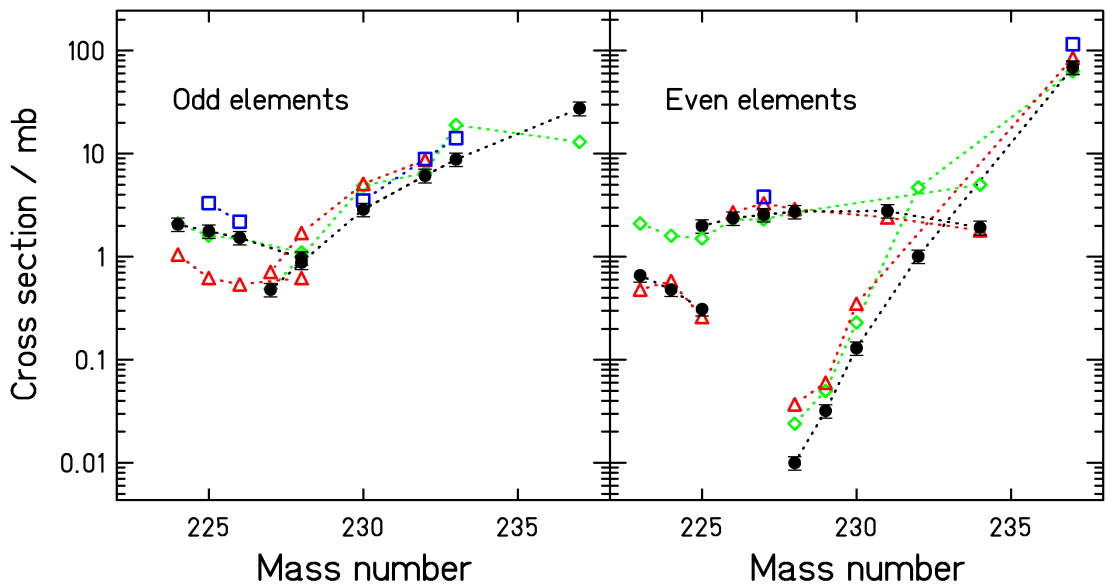

**Figure 7 : Comparison of the cross-sections determined in this work (full points with error bars) with previous results obtained by gamma-spectroscopic methods for $^{237,233,232,230,228,227}$Pa, $^{228,226,225,224}$Ac (left part) and $^{237,232,230,229,228}$U, $^{234,231,228,227,226,225}$Th, $^{225,224,223}$Ra (right part) from refs. [19] (open triangles), [20] (open diamonds), and [21, 22] (open squares). The data points of the same element are connected by dotted lines.**

We observe a systematic disagreement between our data and the ones from Titarenko and collaborators. They systematically overestimate the cross-sections in comparison to ours. On the other hand, many of the data obtained by Lindner and Osborne nicely fit to our results, others, especially the lower cross-sections, deviate by up to a factor of 3 in both directions. Also many data of Pate and Poskanzer agree with our values, while others, in particular those for thorium and uranium isotopes, are considerably higher. Most of the deviations between the different measurements are not systematic. It seems that the older measurements suffer from some normalisation problems for specific elements. In a few cases, we could trace back the origin of the discrepancies to incomplete spectroscopic knowledge at the time of these early publications. Only the data of Titarenko et al. are systematically higher than ours. We have no explanation for this systematic discrepancy. However, we notice that we applied the same procedure than that which was followed for studying the spallation of gold and lead at a similar energy. For those experiments, the normalisation was ensured since the sum of the measured evaporation residues and fission fragment production yielded the total inelastic cross-section, which has been determined with high precision in previous experiments [23]. The investigation of fission residues in our experiment gave a cross-section of 1.53 b [24]. A dedicated measurement of the total fission cross section resulted in (1.52 ± 0.10) b [25]. Together with the estimated total spallation-evaporation-residue cross section of 0.46 b found in this experiment, we obtain a total reaction cross-section of 1.99 b and 1.98 b, respectively. This is in excellent agreement with the calculated total reaction cross section using the Glauber approach described in Karol et al. [26] using updated nuclear-density distributions [27], which results in 1.96 b.

### 3.3 Comparison with systematics

The actual knowledge on the nuclide production in the spallation-evaporation reactions has been compiled by Silberberg, Tsao and Barghouty in an empirical systematics [28]. The parameterisation has recently been updated [29]. The prediction of this systematics is compared with our data in Figure 8. Except for the uranium isotopes, the predictive power of this parameterisation appears relatively low. This is certainly due to the small amount of experimental data available until now. Our data will surely help designing improved analytical expressions aimed at predicting spallation cross-sections.

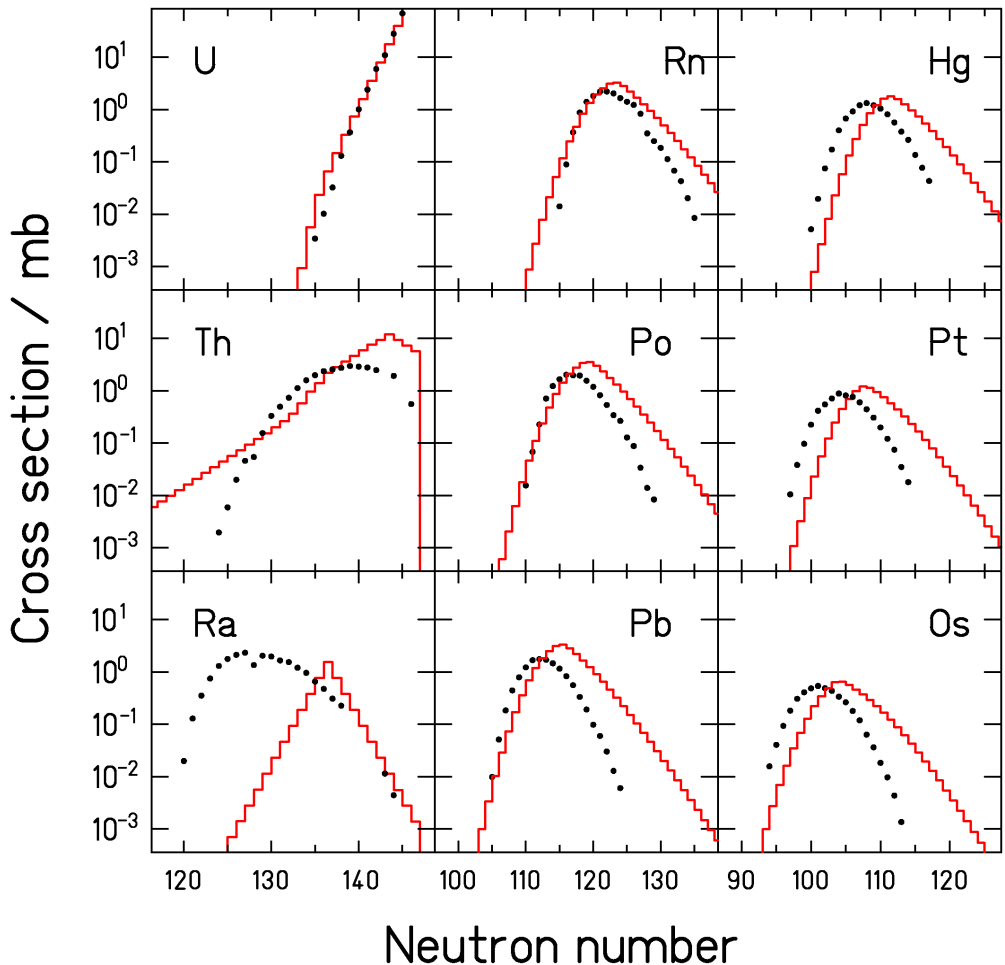

**Figure 8: Comparison of part of the cross-sections determined in this work (full points) with the systematics (histogram) [28,29] . We present the isotopic distribution for even elements.**

## 3.4 Velocity distribution

The experimental set up allows for a measurement of the distributions of the recoil velocities of the produced nuclei. The longitudinal-velocity distributions are well represented by Gaussian distributions. We could determine the mean value and the standard deviation of the recoil-velocity distribution for each ion. We accounted for the slowing down in the layers mounted in the target area, assuming that the nuclear reaction occurred in the middle of the target on the average.

We plotted in Figure 9 the mean velocity normalized following the prescriptions of Morrissey [30]. Thus, we introduce $p'_\parallel$, which is the longitudinal recoil momentum, normalized in the following way:

$$p'_\parallel = v_\parallel \times M_p \times \frac{\beta\gamma}{\gamma+1} \qquad (5)$$

Here, $v_\parallel$ is the velocity of the fragment in the frame of the projectile, and $M_p$, $\beta$ and $\gamma$ are the rest mass and the relativistic parameters of the projectile in the laboratory frame, respectively. This normalisation allows an inter-comparison of various measurements realised at different projectile energies. The location straggling, that means the dependence of the observed velocity on the position of the reaction inside the target, is unfolded for estimating the standard deviation of the velocity distribution. However, this contribution is negligible.

Figure 9 also includes the empirical systematics established by Morrissey [30], which predicts a linear dependence between the reduced recoil momentum ($p'_\parallel$) and the mass loss $\Delta A$ relative to the mass of the projectile. We observe that the systematics describes the measured data reasonably well, although the data points lie above the systematics over the whole mass range.

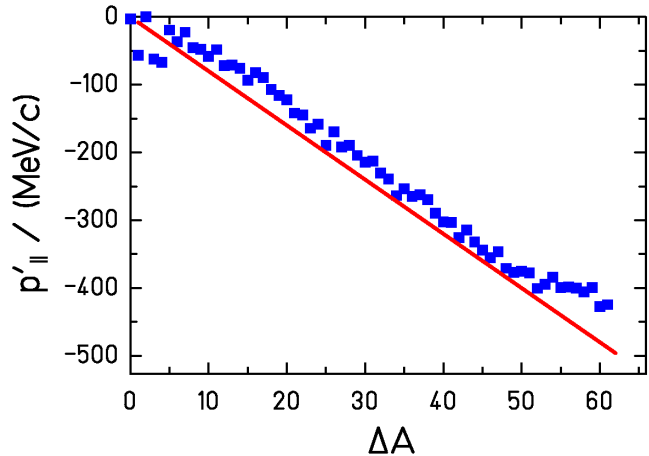

**Figure 9 : Mean recoil momentum induced in the spallation of $^{238}$U by 1 GeV protons as a function of mass loss. The data (symbols) are compared with the systematics of Morrissey (line) [30]. Statistical uncertainties correspond to the scattering of the data. Since the measurement has been performed in inverse kinematics, the measured momenta are transformed into the frame of the beam.**

The width of the longitudinal recoil momentum acquired in the spallation-evaporation reaction is shown in Figure 10. Again, the data are compared with the systematics of Morrissey, and also with the predictions of the Goldhaber model [31]. While the systematics better represents the data for mass losses below $\Delta A = 20$, the experimental values increase more strongly for large mass losses and reach the prediction of the Goldhaber model for $\Delta A \approx 55$. Also in the width of the momentum distribution, isotopic variations are observed which are probably related to the variation of the evaporation contribution to the mass loss [32].

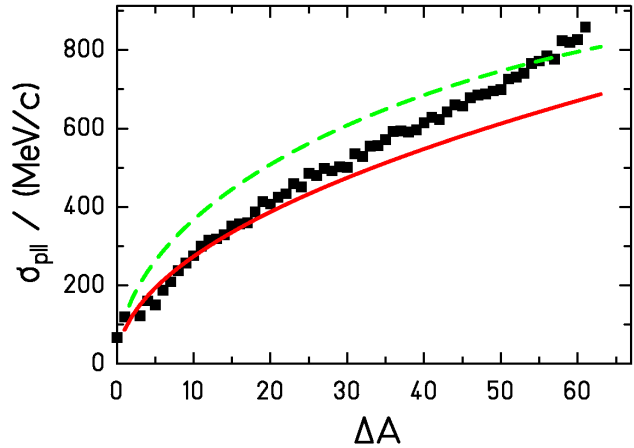

**Figure 10 : Standard deviation of the longitudinal-momentum distribution of the spallation-evaporation residues produced in the bombardment of $^{238}$U with 1 GeV protons. The data (symbols) are compared with the Goldhaber model [31] (dashed line) and with the Morrissey systematics [30] (full line). Since the measurement has been performed in inverse kinematics, the measured momenta are transformed into the frame of the projectiles.**

## 4. Conclusion

The production cross-sections and the longitudinal-momentum distributions of the heavy spallation-evaporation residues from the interaction of 1 $A$ GeV $^{238}$U with hydrogen have been studied, covering elements from tungsten to uranium. The reaction products were fully identified in atomic number $Z$ and mass number $A$ using the magnetic spectrometer FRS.

While the momentum distributions rather well agree with systematics established on the basis of previously measured data, the cross-sections deviate strongly from systematic expectations, especially for neutron-deficient isotopes. These short-lived nuclides could not be measured with alternative techniques previously available.

The data, production cross-sections and kinetic energies, are of highest interest for the design of accelerator-driven systems for the incineration of radioactive waste and as an alternative device for energy production. Using the measured production cross-sections, combined with the known decay properties, the short- and long-term radioactivities in irradiated fissile material can be predicted.

Another field of interest is the production of secondary beams by spallation reactions, a reaction mechanism exploited since many years at ISOLDE and also envisaged for next-generation secondary-beam facilities. The present data give the first comprehensive overview on the reaction cross-sections, and thus provide quantitative information on the secondary-beam intensities potentially available in such facilities, if efficient extraction and ionisation procedures are developed.

The system investigated provides stringent constraints on nuclear-reaction codes, in particular on the modelling of the fission competition. The new data will help to develop improved models with better predictive power for spallation reactions, involving highly fissile nuclei.

**Acknowledgements**
The authors are indebted to K.H. Behr, A. Brünle and K. Burkard for their technical support and assistance during the experiment and to the group of P. Chesny for building the liquid-hydrogen target. This work was supported by the European Union in the frame of HINDAS project and under the programme "Accesss to Research Infrastructure Action of the Improving Human Potential" contract EC-HPRI-CT-1999-00001. One of the authors (J.T) profited from a PhD grant of both IN2P3 and GSI.

accelerator driven transmutation technologies and applications (Prague – Czech Republic), June 1999.